\documentclass[fleqn,twoside]{article}
\usepackage{espcrc2}

\usepackage{graphicx}
\usepackage[figuresright]{rotating}

\newcommand{\AmS}{{\protect\the\textfont2
  A\kern-.1667em\lower.5ex\hbox{M}\kern-.125emS}}
\newcommand{\gapprox}{\raisebox{-0.5ex}{$\
\stackrel{\textstyle>}{\textstyle\sim}\ $}}
\newcommand{\lems}{\hspace{-.65em}}
\newcommand{\lemst}{\hspace{-.6em}}
\title{A BCS Gap on the Lattice}

\author{David N. Walters\address
{Department of Physics, University of Wales
Swansea, \\Singleton Park, Swansea SA2 8PP, United Kingdom.}
        and Simon Hands\addressmark\thanks{PPARC Senior Research Fellow}}

\begin{document}

\begin{abstract}
Monte Carlo simulations of the 3+1 dimensional NJL model are performed
with baryon chemical potential $\mu>0$. For $\mu\gapprox\Sigma_0$, the
constituent quark mass in vacuum, chiral symmetry is restored and a 
diquark condensate $\langle q q_+\rangle$ forms. We analyse 
the fermion propagator and find evidence for particle-hole mixing in the
vicinity of the Fermi surface and an energy gap $\Delta>0$, both
of which provide evidence for superfluidity at high baryon density
induced by a BCS mechanism. 
At $\mu a=0.8$ the ratio between the BCS gap and the vacuum quark
mass is $\Delta/\Sigma_0=0.15(2)$. 
\vspace{1pc}
\end{abstract}

\maketitle

The Nambu -- Jona-Lasinio (NJL) 
model has served for many years as a model of strong
interactions~\cite{Klevansky}; 
in particular with the addition of a non-zero baryon chemical
potential $\mu$ it can be used to model the relativistic quark matter
anticipated in dense systems such as neutron star cores. Phenomena such 
as the BCS-condensation of quark
pairs at the Fermi surface, suggested as a mechanism for color
superconductivity and baryon number superfluidity, 
can be modelled in this way~\cite{RWA}. 
From a lattice theorist's
perspective, the attractive feature of the NJL model is that its action in
Euclidean metric remains real even once $\mu\not=0$, so that standard Monte
Carlo simulation techniques can be employed. Moreover, such simulations 
expose a chiral symmetry restoration transition at a critical
$\mu_c\approx\Sigma_0$, where $\Sigma_0$ is the constituent quark scale, in
qualitative agreement with analytic approaches such as the Hartree
approximation~\cite{Klevansky}.

In previous work~\cite{HW} 
we have simulated the NJL model in an attempt to identify
diquark condensation for $\mu>\mu_c$. In continuum notation the Lagrangian is
\begin{eqnarray}
{\cal L}&\lems=&\lems\bar\psi(\partial{\!\!\!/\,}+m+\mu\gamma_0)\psi\label{eq:L}\\
&\lems-&\lems{\textstyle{g^2\over2}}\Bigl[(\bar\psi\psi)^2
-(\bar\psi\gamma_5\otimes\vec\tau\otimes\gamma_5\psi)^2\Bigr]\nonumber\\
&\lems+&\lems{\textstyle{1\over2}}
\Bigl[
(\bar\psi,\psi^{tr})\left(\matrix{\bar\jmath&\cr&j\cr}\right)
C\gamma_5\otimes\tau_2\otimes C\gamma_5
{\scriptstyle{\left(\matrix{\bar\psi^{tr}\cr\psi\cr}\right)}}\Bigr].
\nonumber
\end{eqnarray}
When formulated using staggered fermions, the resulting lattice model has
$N_f=2$ isospin degrees of freedom, and $N_c=8$ global `colors' (there are no
gluons). The third term in (\ref{eq:L}) is a U(1)$_B$-violating source term
which enables the identification of a superfluid diquark condensate $\langle
q q_+\rangle$ for source strengths
$j=\bar\jmath=j^*\not=0$ on a finite volume.

\vspace{-90ex}
\begin{flushright}
     SWAT/03/355\\
August 2003
\end{flushright}
\vspace{81ex}

At $\mu=0$ the model exhibits breaking of SU(2)$_L\otimes$SU(2)$_R$
into SU(2)$_I$ by the spontaneous generation of a chiral condensate and a
constituent quark mass $\Sigma_0\gg m$. Since in 3+1 dimensions, 
the NJL model is only an effective
theory, the lattice parameters must be chosen to match
low energy phenomenology. Using the large-$N_c$ (Hartree) approximation we
determine a suitable set, leading to physically reasonable results, to be
\begin{equation}
\begin{array}{c c c}
\begin{array}{r c l}
ma&\lemst=&\lemst0.006\\a^{2}g^{-2}&\lemst=&\lemst0.495\\a^{-1}&\lemst=&\lemst720\mbox{MeV}
\end{array}&\lemst\Rightarrow&\lemst
\begin{array}{r c l}
\Sigma_0&\lemst=&\lemst400\mbox{MeV}\\f_\pi&\lemst=&\lemst93\mbox{MeV}\\m_\pi&\lemst=&\lemst138\mbox{MeV}.
\end{array}\end{array}
\label{eq:contnos}
\end{equation}
We use a standard HMC algorithm with a quark kinetic matrix $M$
which is a functional of auxiliary boson fields $\sigma$ and $\vec\pi$.
The source strength $j$ is set to zero in the HMC update, but is allowed to vary
over a range of values in partially
quenched measurements on valence quarks. If we define an antisymmetric matrix
\begin{equation}
{\cal A}={1\over2}\left(\matrix{\bar\jmath\tau_2&M\cr-M^{tr}&j\tau_2\cr}\right),
\end{equation}
then e.g. the diquark condensate is given by
\begin{equation}
\langle q q_+\rangle={1\over{2V}}{{\partial\ln{\cal Z}}\over{\partial j}}=
{1\over{8V}}\langle\mbox{tr}\tau_2{\cal A}^{-1}\rangle.
\end{equation}

The results show that at $\mu_ca\simeq0.6$ there is a transition at
which the chiral condensate decreases sharply until
by $\mu a=0.8$ it has just 5\% of its vacuum value, and at the same point the
baryon density $n_B=(2V)^{-1}\partial\ln{\cal Z}/\partial\mu$ begins to rise
from zero. Results for $\langle q q_+\rangle$, first extrapolated linearly
in $L_t^{-1}$ to $T=0$ from $12^4$, $16^4$ and $20^4$ lattices, 
and then linearly in $j\to0$ using data with $j a\in(0.3,1.0)$, suggest that a
superfluid condensate forms for $\mu a\gapprox0.6$ 
and then increases monotonically
until $\mu a\simeq1.0$, where saturation artifacts become
apparent~\cite{HW}. 

In a system with a Fermi surface, low-energy excitations have 
$k=\vert\vec k\vert$ close to the Fermi momentum $k_F$. 
If a BCS condensate forms,
we expect an energy gap $2\Delta$ to open up between the highest occupied state
in the Fermi sea and the lowest excited state. These excitations are probed
by standard means via temporal decay of
the Euclidean propagator ${\cal G}\equiv{\cal A}^{-1}$. We write 
\begin{equation}
{\cal G}(x,y)=\left(\matrix{A(x,y)&N(x,y)\cr\bar N(x,y)&\bar A(x,y)\cr}\right),
\end{equation}
where we identify both 
``normal'' $N(x,y)\sim\langle q(x)\bar q(y)\rangle$ and
``anomalous'' $A(x,y)\sim\langle q(x)q(y)\rangle$ propagators.
In an isospin-symmetric ground state the only independent components to
survive the quantum average are 
$N\equiv\mbox{Re}N_{11}$ and $A\equiv\mbox{Im}A_{12}$, where subscripts label
isospin components~\cite{HLM}. We proceed by studying the time-slice
propagator ${\cal G}(\vec k,t)=\sum_{\vec x}{\cal G}(\vec0,0;\vec x,t)e^{-i\vec
k.\vec x}$ on a $96\times12^2\times L_t$ lattice ($L_t=12,16,20$ and 24)
with $\vec k=(\pi n/48,0,0)$,
$n=0,1,\ldots,24$. The resulting propagators are fitted with 
\begin{eqnarray}
N(k,t)&=&A e^{-E t}+Be^{-E(L_t-t)}\nonumber\\
A(k,t)&=&C(e^{-E t}-e^{-E(L_t-t)})
\end{eqnarray}
and amplitudes $A,B,C$, and excitation energy $E(k,j)$ extracted.

\begin{figure}[htb]
\vspace{9pt}
\includegraphics[width=17pc]{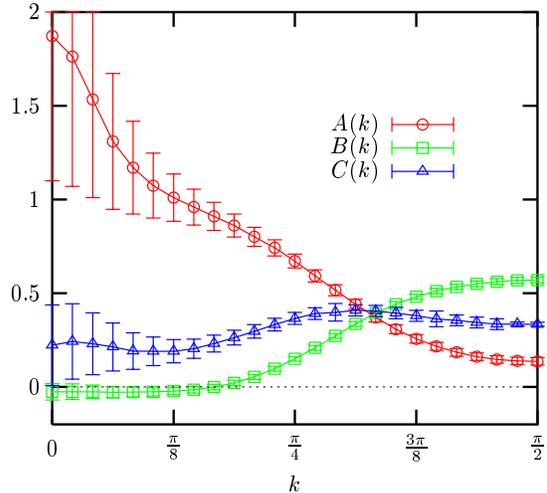}
\caption{Amplitudes at $\mu a=0.8$ extrapolated to $T\to0$ followed by $j\to0$.}
\label{fig:ABC}
\end{figure}
Fig.~\ref{fig:ABC} shows the pole-fit amplitudes as functions of $k$.
For small $k$ the normal propagator is predominantly forward-moving, since
$A\gg B$. For $k\gapprox\mu$, however, the signal becomes predominantly
backward-moving. This is a sign that the excitations for small $k$ are 
hole-like, and for large $k$ particle-like~\cite{HLM}. Of particular interest, 
however, is that in this case even in the limit $j\to0$ there is a region of
$k\approx\mu$
where the anomalous amplitude $C\not=0$. This shows that
there are propagating states which are particle-hole superpositions with
indefinite baryon number, an indirect signal for superfluidity via 
a BCS mechanism.

\begin{figure}[htb]
\includegraphics[width=17pc]{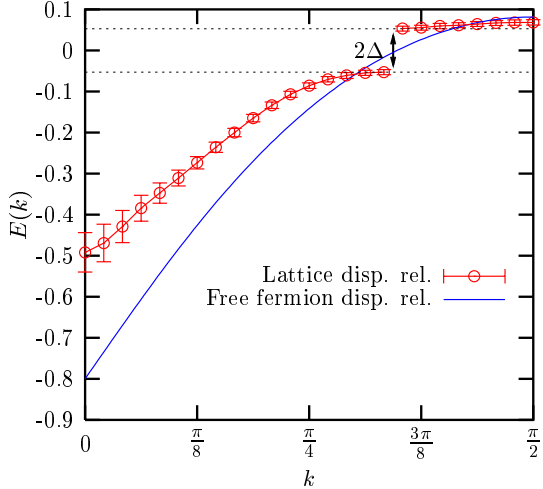}
\caption{Dispersion relation $E(k)$ at $\mu a=0.8$.}
\label{fig:mu0.8}
\end{figure}
In Fig.~\ref{fig:mu0.8} we show the dispersion relation $E(k)$ extracted from
the anomalous propagator at $\mu a=0.8$. It is important to note the order of
extrapolation: first $T\to0$, then $j\to0$. For clarity we have plotted energies
for $k<k_F$ as negative, where the Fermi momentum $k_F$ is chosen to be the
point where $A=B$ in Fig.~\ref{fig:ABC}. We have also plotted the free lattice
fermion dispersion relation, which has $E(\sin^{-1}(\sinh(\mu a))=0$.
The interacting data, however, show no sign of a zero energy state; instead
there is a
a discontinuity at $k\approx k_F$ signalling an energy gap $2\Delta$.
This is the first direct
observation of a BCS gap in a lattice simulation.

\begin{figure}[htb]
\includegraphics[width=17pc]{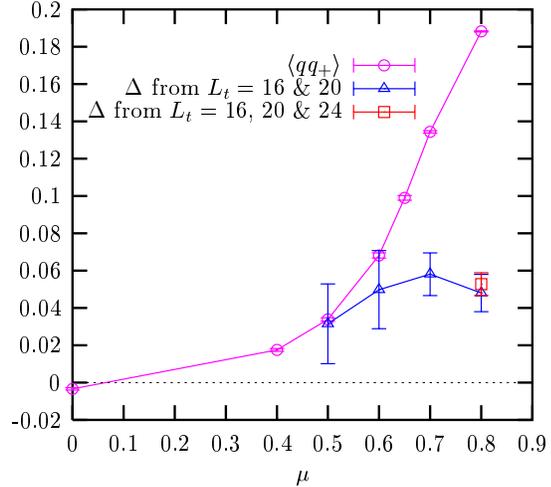}
\caption{$\langle q q_+\rangle$ and $\Delta$ as functions of $\mu$.}
\label{fig:gap}
\end{figure}
Finally in Fig.~\ref{fig:gap} we plot both order parameter $\langle q q_+\rangle$
and gap $\Delta$ as functions of $\mu$. Note that a reliable extrapolation to
$T\to0$ requiring data from $L_t=24$ lattices is only available for $\mu
a=0.8$; simple linear extrapolation through $L_t=16$ and $20$ is completely
consistent with this, however, and hence is used for the other $\mu$ values.
One physical interpretation of the  
order parameter is that it counts the number of $q q$ pairs participating in the
condensate. On the simple-minded assumption that these pairs occupy a
shell of thickness ${\cal O}(\Delta)$ around the Fermi surface, we
surmise the relation  
\begin{equation}
\langle q q_+\rangle\propto\Delta\mu^2.
\end{equation}
Fig.~\ref{fig:gap} indeed suggests that $\Delta$ is roughly constant for
$\mu>\mu_c$, while $\langle q q_+\rangle$ rises markedly. The ratio
$\Delta/\Sigma_0\approx0.15$, which using (\ref{eq:contnos}) translates into
a physical prediction $\Delta\approx60$MeV, consistent with
the model predictions of~\cite{RWA}.

In summary, by presenting numerical estimates for both $\langle q q_+\rangle$ 
and $\Delta$ we have amassed a reasonable body of evidence for 
superfluidity via a 
BCS mechanism in a relativistic quantum field theory, for the first time using a
systematic calculational technique. In future work we hope to address the issue
of finite volume effects, unusually large in this system~\cite{Amore}, which we
believe is due to the difficulty of representing a curved shell of 
states around the Fermi surface on a discrete momentum lattice. It will also be
interesting to study the stability of the superfluid phase for $T>0$, and for 
$\mu_u\not=\mu_d$, which has the effect of separating the Fermi surfaces
of each flavor and hence suppressing isosinglet pairing.


\begin{thebibliography}{9}
\bibitem{Klevansky} S.P. Klevansky, Rev. Mod. Phys. {\bf64} (1992) 649.
\bibitem{RWA} 
K. Rajagopal and F. Wilczek, in: M. Shifman (ed.), {\em Handbook of QCD\/}, 
World Scientific, Singapore, 2001;
M.G. Alford, Ann. Rev. Nucl. Part. Sci. {\bf51} (2001) 131;
M.G. Alford, J.A. Bowers and K. Rajagopal,
J. Phys. {\bf G27} (2001) 235.
\bibitem{HW} S.J. Hands and D.N. Walters, Phys. Lett. {\bf B548} (2002) 196;
D.N. Walters, Nucl. Phys. B (Proc. Suppl.) {\bf119} (2003) 553.
\bibitem{HLM} S.J. Hands, B. Lucini and S.E. Morrison, Phys. Rev. {\bf D65}
(2002) 036004.
\bibitem{Amore} P. Amore, M.C. Birse, J.A. McGovern and N.R. Walet, Phys. Rev.
{\bf D65} (2002) 074005.

\end{thebibliography}
\end{document}